\documentclass[onecolumn]{aastex701}
\usepackage{amsmath}
\usepackage{graphicx}

\usepackage[all]{hypcap}

\makeatletter
\long\def\frontmatter@title@above{}
\def\frontmatter@RRAPrint{}
\makeatother

\begin{document}

\title{TDLight: A Framework for Incremental Light-Curve Management and Continuous Classification Optimization}

\author[orcid=0009-0000-4773-6574]{Xinghang Yu}
\email{3023244355@tju.edu.cn}
\affiliation{School of Computer Science and Technology, Tianjin University, 135 Yaguan Road, Haihe Education Park, Tianjin 300350, People’s Republic of China}
\affiliation{Technical R\&D Innovation Center, National Astronomical Data Center, 135 Yaguan Road, Haihe Education Park, Tianjin 300350, People’s Republic of China}
\author[orcid=0000-0003-2416-4547]{Ce Yu}
\email{yuce@tju.edu.cn}
\affiliation{School of Computer Science and Technology, Tianjin University, 135 Yaguan Road, Haihe Education Park, Tianjin 300350, People’s Republic of China}
\affiliation{Technical R\&D Innovation Center, National Astronomical Data Center, 135 Yaguan Road, Haihe Education Park, Tianjin 300350, People’s Republic of China}
\author[orcid=0009-0005-5107-9796]{Zeguang Shao}
\affiliation{School of Computer Science and Technology, Tianjin University, 135 Yaguan Road, Haihe Education Park, Tianjin 300350, People’s Republic of China}
\affiliation{Technical R\&D Innovation Center, National Astronomical Data Center, 135 Yaguan Road, Haihe Education Park, Tianjin 300350, People’s Republic of China}
\email{shaozeguang@tju.edu.cn}
\author[orcid=0000-0001-9297-0415]{Chen Wang}
\affiliation{College of Computing and Data Science, Nanyang Technological University, 50 Nanyang Ave, Block N 4, Singapore 639798}
\email{chen.wang@ntu.edu.sg}
\author[orcid=0000-0002-3783-2228]{Bin Yang}
\affiliation{School of Computer Science and Technology, Tianjin University, 135 Yaguan Road, Haihe Education Park, Tianjin 300350, People’s Republic of China}
\affiliation{Technical R\&D Innovation Center, National Astronomical Data Center, 135 Yaguan Road, Haihe Education Park, Tianjin 300350, People’s Republic of China}
\email{yangbincic@tju.edu.cn}

\correspondingauthor{Bin Yang}
\email{yangbincic@tju.edu.cn}
\begin{abstract}
   Time-series observations of celestial objects are fundamental to studying variable and transient phenomena. As time-domain surveys continue to scale up, timely automated classification is essential for rapid follow-up. Many existing approaches treat storage and analysis as separate stages, archiving data first and analyzing it later. Yet each source is a continuously growing stream, and early observations often contain sufficient information to assign scientifically useful labels before the full light curve is available. Motivated by this, we present \textsc{TDLight}, a framework for light-curve management and trigger-based classification over incremental data streams. Built on the time-series database \textsc{TDengine}, the system loads the \textit{Gaia} DR2 archive (${\sim}4.8\times10^7$ rows) in 33s and supports low-latency retrieval. As new observations accumulate, \textsc{TDLight} automatically triggers a lightweight LightGBM-based classification pipeline that requires no GPU. When prediction confidence exceeds a predefined threshold, these early results are written back to the database, providing reliable classifications before offline analysis is complete. The deployed system also supports manual label confirmation and optional incremental training. Evaluated on the \textit{LEAVES} dataset (${\sim}1.10$ million sources spanning 10 classes), the system achieves an overall accuracy of 93.6\% and a weighted F1 score of 93.7\%. Furthermore, it identifies 560 high-confidence catalog-disagreement candidates for follow-up verification. Notably, this includes correcting 150 RRc variables erroneously labeled as EW binaries due to inherited period-doubling artifacts. \textsc{TDLight} and the related data products are publicly available.
\end{abstract}


\section{Introduction} \label{sec:intro}

Time-domain astronomy studies the temporal evolution of celestial objects through their light curves, which encode critical information about phenomena ranging from stellar pulsations to explosive transients. The identification and classification of variable sources from these light curves is fundamental to modern astrophysics. Certain classes of variable stars, particularly Cepheids and RR~Lyrae stars, serve as important distance indicators and provide key constraints on stellar structure and evolution \citep{riess20162, clementini2019gaia}. Observations of Type~Ia supernovae provided key evidence for the accelerated expansion of the Universe \citep{riess1998observational, perlmutter1999supernovae}, and the discovery of electromagnetic counterparts to gravitational waves has underscored the need for rapid analysis and follow-up \citep{abbott2017multi}. Across all these domains, scientific yield is maximized when targets are accurately characterized at early observational stages, making timely and automated classification a central requirement.

The landscape of time-domain surveys has scaled dramatically, from early projects such as MACHO \citep{alcock2000macho} and OGLE \citep{udalski2015ogle} to modern high-cadence facilities including the Zwicky Transient Facility \citep[ZTF;][]{bellm2019ztf} and \textit{Gaia} \citep{gaia2023dr3}. The next-generation Vera C.\ Rubin Observatory Legacy Survey of Space and Time (LSST) will extend this trend further, making automated classification and anomaly detection indispensable \citep{ivezic2019lsst, fluke2020surveying}. Accordingly, a broad range of machine-learning methods has been developed for light-curve analysis, from traditional feature-based classifiers to recent deep-learning models, including foundation-model approaches \citep{yu2021light, richards2011mlvar, kim2014epoch, disanto2016feature, zuo2026falco}. In practice, however, these methods are still most often applied in an offline manner after data collection, rather than being embedded into the data-management workflow. This separation is increasingly limiting for incrementally growing light curves, whose classifications may need to be revisited as new observations arrive. Moreover, the computational and engineering demands of many existing models make frequent, low-latency re-analysis difficult in lightweight local deployments. As a result, classification is commonly treated as a post-processing task, decoupled from the primary storage and update loop.

Existing approaches fall into two broad categories. \emph{Archival systems} are optimized for time-series storage and retrieval: for example, \textsc{LCGCT} \citep{zhang2024lcgct} uses a purpose-built time-series database to replace a traditional relational backend, achieving substantial storage savings and faster photometric queries; however, scientific analysis in such systems is still typically performed after data retrieval. \emph{Real-time alert brokers}, such as ALeRCE \citep{forster2021alerce} and the broader ZTF alert infrastructure \citep{patterson2019ztfalerts, masci2019ztfarchive}, are designed for community-scale alert distribution and typically separate real-time event handling from long-term source-centered local storage. In practice, many light curves are formed progressively as catalogues are cross-matched over time, and neither category supports repeated source-level analysis during data accumulation. What is still lacking is a lightweight database-centered framework that supports repeated source-level classification and confidence-controlled write-back as new observations accumulate.

\begin{figure}[t]
\centering
\includegraphics[width=\textwidth]{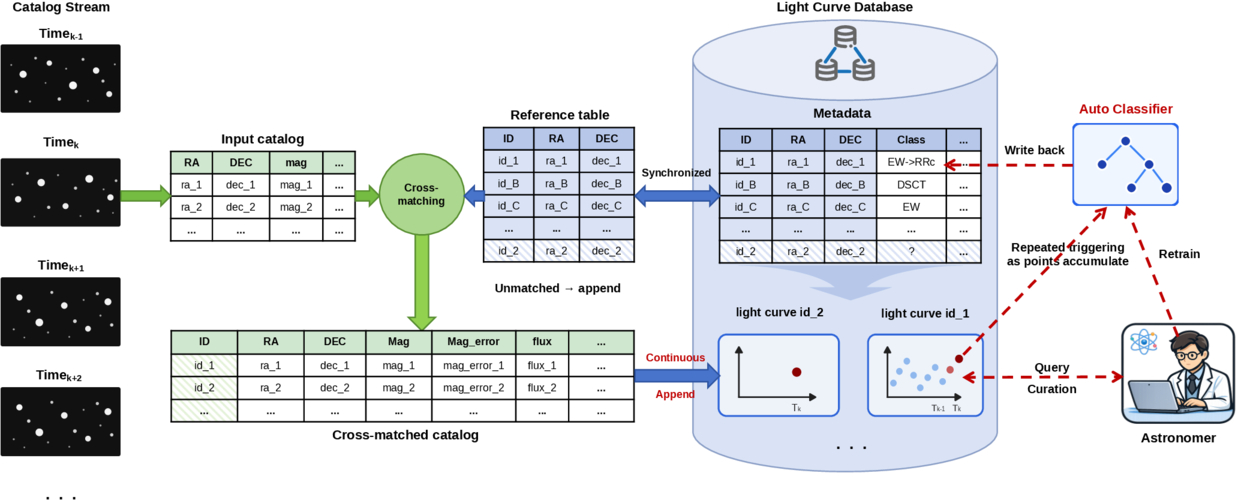}
\caption{Overview of the \textsc{TDLight} workflow for incrementally arriving catalogue streams. New catalogues are cross-matched against a reference table and appended to a source-centered light-curve database. As each source accumulates new observations, the system re-triggers feature extraction and classification, and writes high-confidence results back to the database before the full light curve becomes available.}
\label{fig:paper_diagram}
\end{figure}

We present \textsc{TDLight}, a framework that unifies ingestion, reclassification, and database write-back within a single source-centered workflow, embedding classification directly into the data path rather than treating it as a separate offline task. Built on \textsc{TDengine}\footnote{\url{https://www.tdengine.com}},
the system manages each source as an independent time-series entity with HEALPix spatial indexing, supporting high-throughput incremental ingestion and sub-second retrieval. When sufficient new observations have accumulated, \textsc{TDLight} automatically re-extracts features and runs a LightGBM-based hierarchical classifier, writing results back to the database only when the prediction confidence exceeds a predefined threshold and deferring uncertain cases until more data arrive. As evidence progressively accumulates, the features become increasingly representative, allowing reliable results to be accepted before the full light curve is available.

Beyond its software architecture, \textsc{TDLight} serves as a powerful tool for catalog auditing and data curation. By deploying the framework on the multi-survey \textit{LEAVES} dataset, we identified 560 high-confidence misclassifications in existing archives. Most notably, we discovered and corrected 150 RRc variables that had been systematically misclassified as W Ursae Majoris (EW) binaries due to inherited period-doubling artifacts. Along with the lightweight, GPU-free open-source software, we release a curated supplementary catalog of these corrected sources to the community.

This paper is organized as follows. Section~\ref{sec:data} describes the two datasets used in this study. Section~\ref{sec:framework} introduces the overall \textsc{TDLight} framework, including the system architecture, data model, and data flow. Section~\ref{sec:experiments} evaluates the framework in terms of incremental ingestion, spatial query efficiency, hierarchical classification performance on \textsc{LEAVES}, early classification enabled by trigger-based updates, and the discovery of systematic catalog disagreements.
Section~\ref{sec:deployment} describes the deployment environment and web interface. Section~\ref{sec:conclusion} concludes with a summary and prospects for future work.

\section{Data Preparation} \label{sec:data}

We use two datasets with complementary roles in the experiments. The \textit{Gaia}~DR2 variable-star archive provides the large-scale time-series workload for ingestion and retrieval benchmarks, while the \textsc{LEAVES} catalog provides labeled samples for classification evaluation. Table~\ref{tab:datasets} summarizes both datasets.

\begin{table}[ht]
\caption{Summary of the two datasets used in this work.\label{tab:datasets}}
\centering
\begin{tabular}{lcc}
\hline\hline
 & \textit{Gaia} DR2 & \textsc{LEAVES} \\
\hline
Sources              & 549\,370                & ${\sim}$1.10\,M \\
Photometric records  & ${\sim}4.8\times10^{7}$ & ${\sim}3.8\times10^{8}$ \\
Classes              & ---                     & 10 leaf classes \\
Role                 & Ingestion \& query benchmark & Classification training \& evaluation \\
\hline
\end{tabular}
\end{table}

\subsection{Gaia DR2 Light-Curve Archive}

We retrieved epoch photometry for all 549,370 confirmed variable sources from
the \textit{Gaia} Data Release 2 (DR2) variability tables, using the public
\textit{Gaia} archive \citep{gaia2018dr2,holl2018dr2var}. The Gaia DR2 data
used in this work are available from the ESA Gaia Archive\citep{giga2018}.
The resulting dataset contains approximately $4.8\times10^7$ individual photometric measurements in the $G$, $G_{\rm BP}$, and $G_{\rm RP}$ bands, whose photometric content and calibration are described by \citet{evans2018dr2phot}. Because the variability tables do not directly provide the astrometric positions required by our storage and spatial-query pipeline, we joined them with the main \textit{Gaia} DR2 source catalog through \texttt{source\_id} to obtain right ascension and declination, and then converted these coordinates into HEALPix indices ($N_{\rm side}=64$) for spatial indexing (Section~3).

We adopted DR2 in this work because it already provides a sufficiently large and realistic benchmark for evaluating the ingestion and indexing behavior of TDLight under large-scale workloads. In addition, the DR2 data can be reorganized in a straightforward and reproducible way into the source-level and epoch-level layouts required by our experiments. While \textit{Gaia} DR3 provides expanded variability products, including epoch photometry for many more sources, these data are distributed through the archive's DataLink products rather than simple table queries, and incorporating them into the same benchmark would require additional data preparation. We therefore leave extension to later \textit{Gaia} releases to future work.

To evaluate both ingestion modes of TDLight, we reorganized the raw data into two complementary layouts: (i) 549,370 per-source CSV files, each containing the complete multi-band light curve of a single object, for the archival (batch) benchmark; and (ii) 728 epoch-style catalog files, each containing a single-epoch snapshot of many sources, to simulate streaming alert ingestion.

\subsection{LEAVES Classification Dataset} \label{subsec:data_leaves}

For classification, we adopt the \textsc{LEAVES} dataset
\citep{fei2024leaves}, a multi-survey variability catalog drawn from
ASAS-SN, ZTF, and \textit{Gaia}. The \textsc{LEAVES} dataset used in this work is available \citep{leaves_data}.
The underlying survey resources include ASAS-SN light curves
\citep{kochanek2017asas,jayasinghe2018asas,
jayasinghe2019asas,christy2023asas}, ZTF light curves hosted by IPAC \citep{irsa598} 
and \textit{Gaia} DR3 data products \citep{gaia2022}.
The dataset contains ${\sim}$1.10~million labeled sources organized into
10 leaf classes under a four-level hierarchical taxonomy.
Table~\ref{tab:leaves_classes} lists the per-class counts; EW and SR
together account for over 60\% of the sample, while CEP is the rarest
class. We use the published labels and light curves without modification.

\begin{table}[ht]
\caption{Class distribution of the \textsc{LEAVES} 10-class dataset.\label{tab:leaves_classes}}
\centering
\begin{tabular}{lrlr}
\hline\hline
Class & Count & Class & Count \\
\hline
EW      & 336\,875 & RRAB  & 41\,823 \\
SR      & 331\,111 & M     & 19\,168 \\
Non-var & 134\,592 & RRc   & 18\,703 \\
ROT     & 119\,112 & DSCT  & 18\,600 \\
EA      &  77\,356 & CEP   &  2\,455 \\
\hline
\end{tabular}
\end{table}

For each source, we extract 15 statistical features from its light curve using the \textsc{feets} package \citep{cabral2018feets}, including the Lomb--Scargle period \citep{lomb1976least, scargle1982studies}, Stetson~$K$ index, amplitude, and skewness. These features are used as input to the hierarchical LightGBM classifier described in Section~\ref{subsec:classification}.

\section{TDLight Framework} \label{sec:framework}

\subsection{Architectural Overview}
The \textsc{TDLight} framework builds on \textsc{TDengine}, a database originally designed for Internet of Things (IoT) telemetry, and adapts it to astronomical time-domain data. Because light curves are inherently time-ordered, a time-series database is a natural backend for their storage and incremental management. In this design, each astronomical source is treated as an independent time-series entity, while \textsc{TDengine}'s columnar architecture and native support for time-series workloads provide an efficient basis for append-oriented ingestion, compression, and source-centered data management.

\textbf{Data model.} \textsc{TDLight} adopts a ``one-table-per-source'' schema, in which each source is stored as an individual \textsc{TDengine} \emph{child table} beneath a shared \emph{supertable}. This layout matches both \textsc{TDengine}'s native data model and the source-centered nature of light-curve data. Observations of the same source remain grouped in time order, while static attributes such as coordinates, HEALPix indices, and class tags are stored once as tag columns, making the design well suited to incremental ingestion and per-source retrieval.

For comparison, a conventional relational design stores all sources in a single shared table indexed by \texttt{source\_id} and timestamp. Although flexible, this layout is less naturally matched to the source-centered incremental workflow of \textsc{TDLight}, in which observations are appended over time and classification results are repeatedly updated at the source level. As shown in Table~\ref{tab:db_compare}, the per-source schema provides higher ingestion throughput and lower storage overhead than the relational baselines under this source-centered workload. Overall, this behavior is consistent with the design goals of \textsc{TDLight}.

\begin{figure}[t]
\centering
\includegraphics[width=\textwidth]{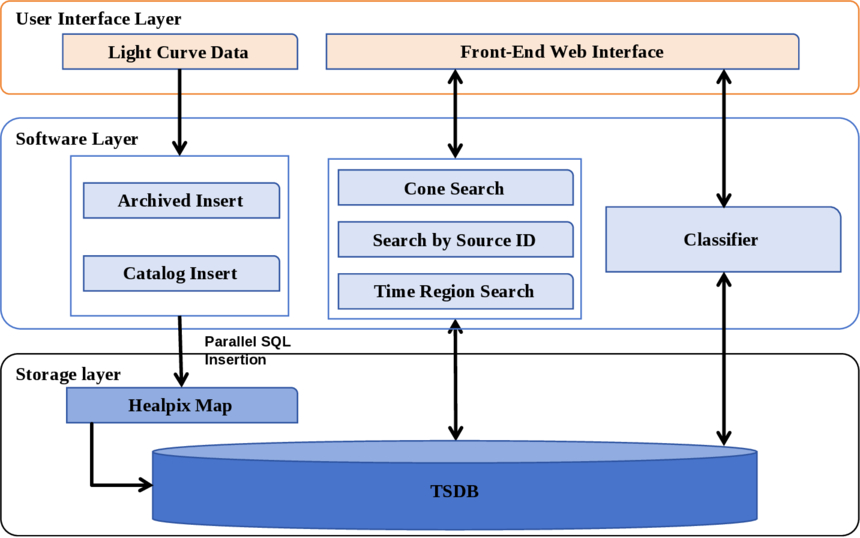}
\caption{Overview of the \textsc{TDLight} framework, including the user interface, software, and storage layers.}
\label{fig:framework}
\end{figure}

Figure~\ref{fig:framework} summarizes the overall architecture of \textsc{TDLight}, which is organized into three coordinated layers: storage, software, and user interface.

The \textbf{Storage Layer} manages both photometric time-series data and static source metadata in \textsc{TDengine}. Time-varying observations are stored in per-source child tables, while coordinates, HEALPix indices, and classification tags are recorded as tag columns.

The \textbf{Software Layer} contains three main modules: data ingestion, data retrieval, and classification. The ingestion module parses external files and inserts records into the database; the retrieval module translates user queries into source-level and spatial database operations; and the classification module is triggered after data updates, performing feature extraction, LightGBM-based inference, and high-confidence result write-back.

The \textbf{User Interface Layer} is a web-based frontend through which users inspect light curves, query sources, and view classification results. It communicates with the Software Layer through REST APIs rather than accessing the database directly.

\subsection{Data Flow}
\textsc{TDLight} supports two complementary ingestion modes. In archival mode, as in the \textit{Gaia}~DR2 benchmark, each source is stored as an individual CSV file and ingested into its corresponding per-source table. In streaming mode, each incoming catalog contains a single-epoch snapshot of many sources. To maintain source continuity across epochs, these records are first associated with existing sources through a reference table, following the reference-table-based source association idea of \textsc{AstroCatR} \citep{yu2020astrocatr} and the broader literature on astronomical cross-identification \citep{budavari2008crossid}. The reference table stores source identifiers together with positional metadata and is used—through HEALPix-based candidate filtering followed by sky-position matching—to determine whether an incoming record should be matched to an existing source or initialized as a new one. Matched records are appended to the corresponding per-source tables, while unmatched records are inserted with newly created source metadata. In both modes, HEALPix indices are computed from the source coordinates and recorded as metadata for subsequent spatial queries.

After new observations are inserted, the system schedules feature extraction and LightGBM-based classification for newly ingested sources as well as previously seen sources whose light curves have grown sufficiently since the last update. When the prediction confidence exceeds a predefined threshold, the classification result is written back to the database as a tag update; otherwise, the source remains unlabeled until more observations become available.

Users can interactively query the database through the web frontend using cone searches, source ID lookups, or rectangular sky regions, and inspect both the accumulated light curves and the current classification results. In this way, ingestion, cross-matching, storage, repeated classification, and result write-back are integrated into the same workflow.

\section{Experiments and Evaluation} \label{sec:experiments}

We benchmarked system performance using a dataset derived from \textit{Gaia} DR2, comprising 549,370 unique objects and $\sim$48.2 million photometric records. The data were reorganized into two formats: (i)~549,370 individual per-source CSV files for archival loading, and (ii)~728 epoch-style catalog files---each containing a single-epoch snapshot of multiple sources---for simulated alert-stream ingestion. All benchmarks were conducted on a dual-socket Intel Xeon Gold 6530 server (64~threads, NVMe SSD) running \textsc{TDengine}~3.4.

\subsection{Ingestion}

\begin{figure}[!hbtp]
\centering
\includegraphics[width=\textwidth]{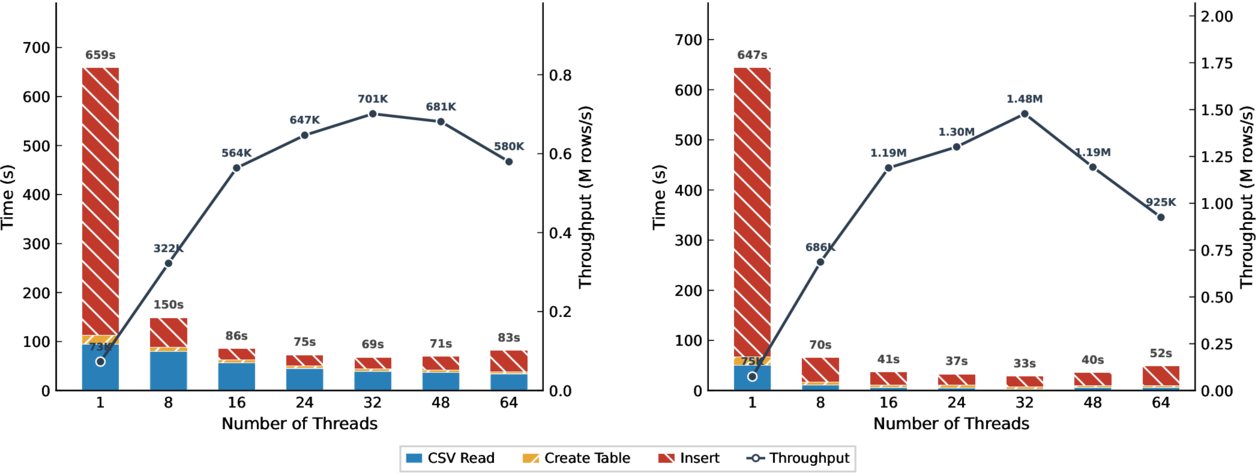}
\caption{Ingestion performance on the \textit{Gaia} DR2 benchmark. Left: streaming mode, peaking at $7.0 \times 10^{5}$~rows~s$^{-1}$. Right: archival mode, peaking at $1.48 \times 10^{6}$~rows~s$^{-1}$.}
\label{fig:ingestion}
\end{figure}

\textbf{Archival Ingestion (Batch Mode).} In this mode, 549,370 per-source light curve files are loaded in parallel. Source coordinates are precomputed into HEALPix indices ($N_{\rm side}=64$) and cached locally. Concurrent worker threads execute batched insertions, leveraging \textsc{TDengine}'s automatic table creation to eliminate pre-scanning overhead. As shown in the right panel of Figure~\ref{fig:ingestion}, this pipeline achieves a peak end-to-end throughput of $1.48 \times 10^{6}$~rows~s$^{-1}$ at 32~threads, completing the full 48.2-million-row ingest in 33~s.

\textbf{Streaming Ingestion (Real-time Mode).} To simulate alert streams, epoch-style catalogs are ingested sequentially: each catalog is parsed, and individual records are routed to the corresponding per-source tables via HEALPix mapping. To minimize database round-trips, the ingestion engine aggregates insertion statements targeting multiple child tables into unified batch requests. The left panel of Figure~\ref{fig:ingestion} shows that this strategy sustains a peak throughput of $7.0 \times 10^{5}$~rows~s$^{-1}$ at 32~threads.

Both modes exhibit a throughput peak near 32~threads followed by a slight decrease at higher thread counts, which we attribute to lock contention within the storage engine when the number of concurrent writers exceeds the number of physical cores per socket.

\textbf{Comparison with alternative database backends.}
Figure~\ref{fig:ingestion} shows the end-to-end ingestion performance of \textsc{TDLight} on the full archival benchmark described above. For comparison, we benchmarked the same framework with three relational backends---PostgreSQL\footnote{\url{https://www.postgresql.org/docs/}}, TimescaleDB\footnote{\url{https://docs.timescale.com/self-hosted/latest/}}, and SQLite\footnote{\url{https://www.sqlite.org/}}, using the same benchmark workload, a unified C++17 implementation, and identical CSV parsing, HEALPix computation, and timing logic. For fairness, \textsc{TDengine}, PostgreSQL, and TimescaleDB were driven by 32 client threads, whereas SQLite was evaluated with a single writer thread because its write path remains globally serialized and additional threads degraded performance. Table~\ref{tab:db_compare} summarizes the comparison.

With the \textsc{TDengine} backend, the framework achieves the highest ingestion throughput, reaching $1.63\times10^6$ rows~s$^{-1}$, compared with $6.98\times10^5$ rows~s$^{-1}$ for PostgreSQL, $3.43\times10^5$ rows~s$^{-1}$ for TimescaleDB, and $1.10\times10^6$ rows~s$^{-1}$ for SQLite. It also yields the smallest on-disk footprint (4.4\,GB, versus 8.5\,GB for PostgreSQL and TimescaleDB, and 6.8\,GB for SQLite). PostgreSQL and TimescaleDB show broadly similar behavior in this workload, with TimescaleDB incurring a noticeable reduction in ingestion throughput because its hypertable chunk routing adds overhead during bulk insertion. SQLite achieves surprisingly high throughput despite running with a single writer thread, reflecting its low embedded overhead; however, it is included primarily as a lightweight local baseline rather than as a head-to-head competitor to the server-grade backends. Overall, these results indicate that the per-source schema enabled by \textsc{TDengine} is well suited to high-throughput archival ingestion under the source-centered workload of \textsc{TDLight}, while the relational baselines provide lower write performance and larger storage footprints under the same schema constraints. This combination of higher ingestion throughput and a more compact archive footprint is particularly attractive for downstream light-curve services in survey environments approaching the scale of Rubin Observatory's Legacy Survey of Space and Time (LSST; \citealt{ivezic2019lsst}). Current Rubin project materials indicate a nightly data volume of about 10~TB and an alert stream of order $10^7$ events per night.\footnote{\url{https://rubinobservatory.org/for-scientists/rubin-101/key-numbers}} Although our benchmark addresses source-level photometric ingestion rather than raw image handling, the measured results suggest that \textsc{TDLight} can serve as an efficient archive layer for such time-domain workloads.

\begin{table}[ht]
\caption{Archival-mode performance with per-source and relational schemas. All results are obtained from the same C++17 benchmark framework on the same server. Throughput is computed from pure database time excluding CSV parsing. For the relational backends, no secondary indexes are created on the single-table schema; SQLite is evaluated with one writer thread because concurrent writes are serialized by its locking model, whereas PostgreSQL, TimescaleDB, and \textsc{TDLight} use 32 threads. Cone-search latencies are measured with 500 random queries (0.5\degr\ radius) that perform direct \textit{ra/dec} spatial filtering via SQL.\label{tab:db_compare}}
\centering
\makebox[\columnwidth][c]{\begin{tabular}{lcccc}
\hline\hline
 & SQLite & PostgreSQL & TimescaleDB & \textsc{TDLight} \\
\hline
Threads                           & 1      & 32     & 32     & 32 \\
Write Throughput (rows\,s$^{-1}$) & 1.10\,M & 698\,k & 343\,k & 1.63\,M \\
Cone-search med.\ (ms)            & 2152   & 1089   & 1070   & 86.44 \\
Cone-search P90 (ms)              & 2383   & 1172   & 1300   & 131.95 \\
Cone-search P99 (ms)              & 2464   & 1215   & 1858   & 337.06 \\
Disk (GB)                         & 6.8    & 8.5    & 8.5    & 4.4 \\
\hline
\end{tabular}}
\end{table}

\subsection{Spatial Query}

For cone searches, \textsc{TDLight} adopts a HEALPix-based filter-and-refine strategy. HEALPix-based metadata filtering is first used to restrict the search to a small set of relevant sky pixels, after which an in-memory geometric refinement is applied to the remaining candidates. In the \textsc{TDLight} implementation, this strategy reduces cone-search latency from about 10\,s for a full-table scan to 0.10--0.38\,s over search radii from 1\arcmin\ to 20\arcmin, corresponding to speedups of approximately $27\times$ to $104\times$.

To assess cross-backend cone-search performance under identical spatial semantics, we benchmarked 500 random queries (0.5\degr\ radius) using a unified C++17 framework. All backends execute the same cone-search logic: a rectangular $ra/dec$ bounding-box pre-filter followed by an exact angular-distance test.

Table~\ref{tab:db_compare} summarizes the results. Under this spatial-filter workload, the \textsc{TDLight} storage layout achieves a median latency of 86\,ms, compared with 1.07\,s for TimescaleDB, 1.09\,s for PostgreSQL, and 2.15\,s for SQLite. The difference reflects the design-level mismatch between the source-centered spatial-query pattern and the conventional single-table relational schema: \textsc{TDLight} resolves each cone to a small set of HEALPix pixels and queries only the corresponding per-source subtables, whereas the relational backends must evaluate the same spatial predicate across the entire monolithic table without a spatial index. TimescaleDB and PostgreSQL show broadly similar median latency, with TimescaleDB exhibiting a larger tail (P99 = 1.86\,s versus 1.22\,s for PostgreSQL) because its hypertable chunk routing adds overhead without benefit when no time-based or spatial pruning is possible. The SQLite result (2.15\,s median) is included as a lightweight embedded reference rather than as a head-to-head competitor.

These results show that, under the source-centered workload of \textsc{TDLight}---where spatial queries are central and the data model is designed around per-source tables---the HEALPix-based per-source layout outperforms conventional single-table relational baselines by a factor of $12\times$--$25\times$ in cone-search latency, while also retaining the higher write throughput and lower storage footprint reported above.

\subsection{Classification} \label{subsec:classification}

\begin{figure}[t]
\centering
\includegraphics[width=0.75\textwidth]{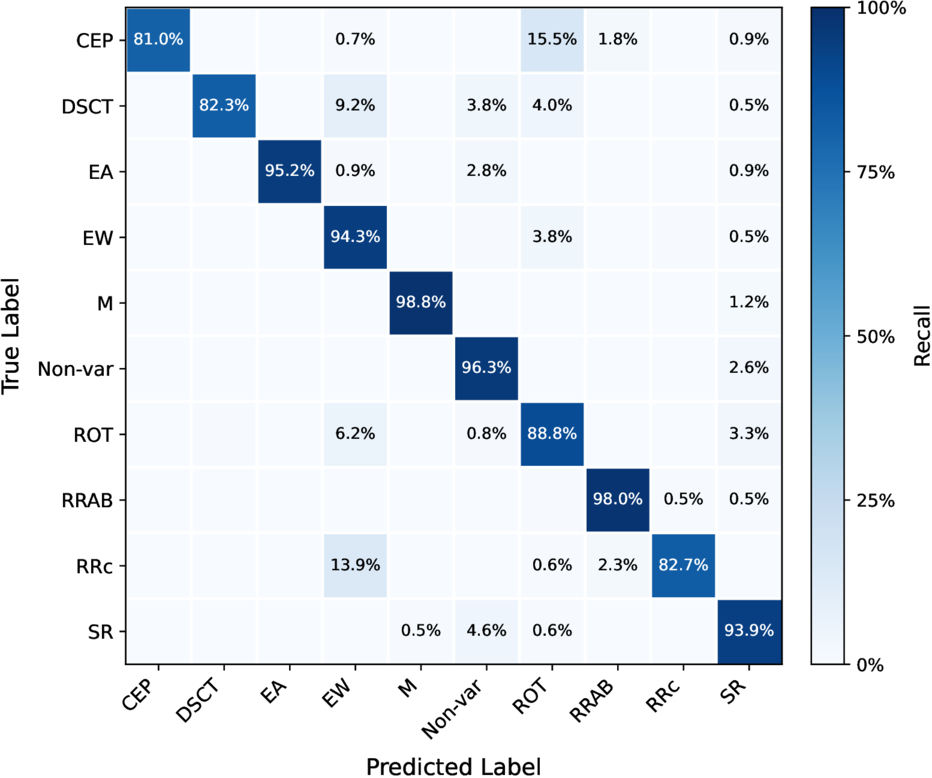}
\caption{Normalized confusion matrix from 5-fold cross-validation on the \textsc{LEAVES} dataset (1.10\,M sources, 10 classes). Diagonal entries denote per-class recall; off-diagonal entries indicate the dominant misclassification channels. The overall accuracy is 93.6\%.}
\label{fig:confusion}
\end{figure}

For classification, \textsc{TDLight} uses the \textsc{LEAVES} dataset \citep{fei2024leaves}, a publicly released multi-survey light-curve dataset constructed from ASAS-SN, ZTF, and \textit{Gaia}, containing ${\sim}1.10$~million sources in 10 leaf classes. The original \textsc{LEAVES} balanced-random-forest (BRF) models achieved a weighted F1 score of 0.93 on the 10-class task but are too heavy for lightweight online deployment: their serialized files total 21.8\,GB and require more than 22\,GB of resident memory at runtime (Table~\ref{tab:inference}), with a loading time of about 18\,s. We therefore retain the same four-level hierarchical architecture but replace the BRF implementation with LightGBM \citep{ke2017lightgbm}, yielding a much more compact classifier for CPU-only deployment. The classification module is designed to be replaceable; the interface requirements are provided in Appendix~\ref{app:classification-interface}.

\textbf{Model architecture.} The classifier retains the same four-level decision tree used by \textsc{LEAVES}: (1)~Variable vs.\ Non-variable, (2)~Extrinsic vs.\ Intrinsic, (3)~subclass routing (e.g.\ EB vs.\ ROT; RR vs.\ LPV vs.\ CEP vs.\ DSCT), and (4)~fine-grained leaf classifiers (e.g.\ RRAB vs.\ RRc; EA vs.\ EW; Mira vs.\ SR). This yields seven individual LightGBM sub-models, each operating on 15 statistical features extracted by the \textsc{feets} package \citep{cabral2018feets}, including the Lomb--Scargle period, Stetson~K index, amplitude ratios, and variability measures. For production deployment, the trained models are further exported to ONNX format.

\textbf{Training and accuracy.} All seven sub-models were trained on the full \textsc{LEAVES} dataset and evaluated via stratified 5-fold cross-validation. The resulting ensemble achieves an overall accuracy of 93.6\% and a weighted F1 score of 93.7\% on the 10-class task (Figure~\ref{fig:confusion}). For comparison, the original \textsc{LEAVES} hierarchical RF baseline reports a weighted F1 score of 0.93 on the same 10-class setting. The difference in the overall weighted metric is modest. The original baseline shows stronger class-dependent variation, with particularly weak results for minority subclasses such as DSCT (F1 = 0.80) and low precision for CEP (0.62); our model exhibits a more even recall profile across classes. In our model, per-class recall ranges from 81.0\% (CEP) to 98.8\% (M), with the dominant classes (EW, SR) both exceeding 93\%.

\textbf{Deployment efficiency.} As shown in Table~\ref{tab:inference}, the LightGBM ensemble reduces the total model footprint from 21.8\,GB to 252\,MB---an $87\times$ reduction---while requiring less than 10\,MB of runtime memory. Although the pure inference time is slightly higher than that of the original BRF model (17.6\,s versus 14.3\,s), the dominant bottleneck in practice is model loading rather than tree traversal, and the loading time drops from 18.2\,s to 4.5\,s. As a result, the end-to-end latency is reduced from 32.4\,s to 22.2\,s. Moreover, \textsc{TDLight} is designed for incremental and on-demand classification during database updates, rather than for repeatedly loading and classifying the entire \textsc{LEAVES} corpus at once. Under this deployment pattern, the much smaller model footprint is more important than the modest increase in per-batch inference time. The resulting system runs on a single server without GPU support.

\begin{deluxetable}{lcc}
\tablecaption{Inference comparison: LEAVES BRF vs.\ LightGBM. Full dataset (${\sim}1.10$\,M sources), 64 CPU threads, Intel Xeon Gold 6530. \label{tab:inference}}
\tablehead{
    \colhead{Metric} & \colhead{LEAVES BRF} & \colhead{LightGBM}
}
\startdata
Model size (disk)     & 21.8\,GB            & 252\,MB ($87\times$) \\
Memory (RSS)          & 22.6\,GB            & $<$10\,MB \\
Load time             & 18.2\,s             & 4.5\,s ($4.0\times$) \\
Inference time        & 14.3\,s             & 17.6\,s \\
End-to-end latency    & 32.4\,s             & 22.2\,s ($1.5\times$) \\
\enddata
\end{deluxetable}

\subsection{Early Classification and Confidence Filtering} \label{subsec:early}

\begin{figure}[t]
\centering
\includegraphics[width=\textwidth]{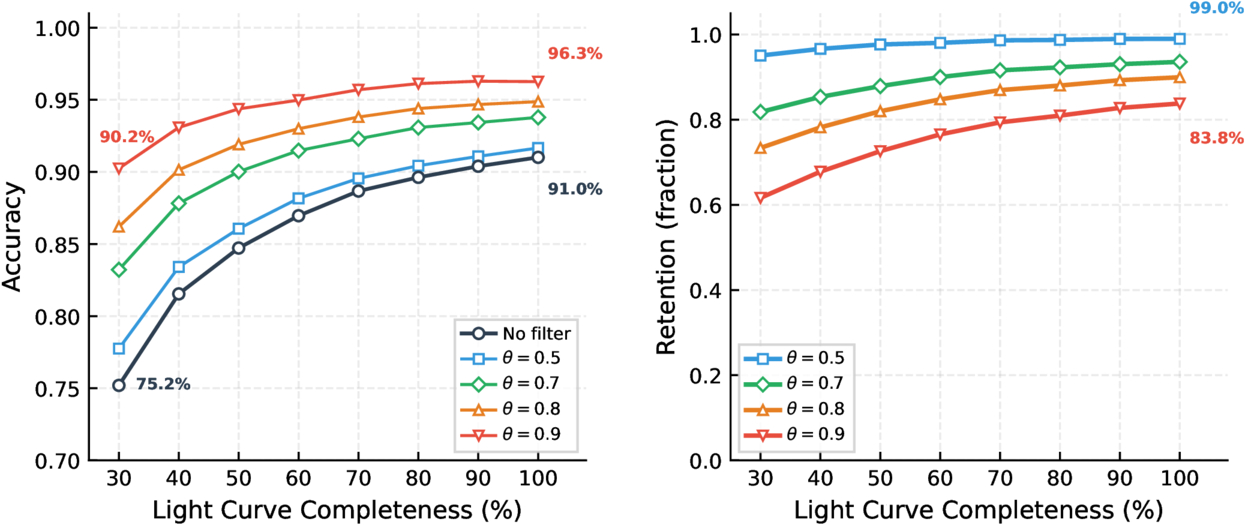}
\caption{Confidence filtering vs.\ light-curve completeness. \textbf{(a)}~Classification accuracy without filtering and with filtering on predictions satisfying $c \geq \theta$ for several $\theta$. \textbf{(b)}~Overall retention fraction (fraction of sources with $c \geq \theta$) for the same thresholds; stricter $\theta$ lowers retention.}
\label{fig:cv_threshold}
\end{figure}

\begin{figure}[t]
\centering
\includegraphics[width=\textwidth]{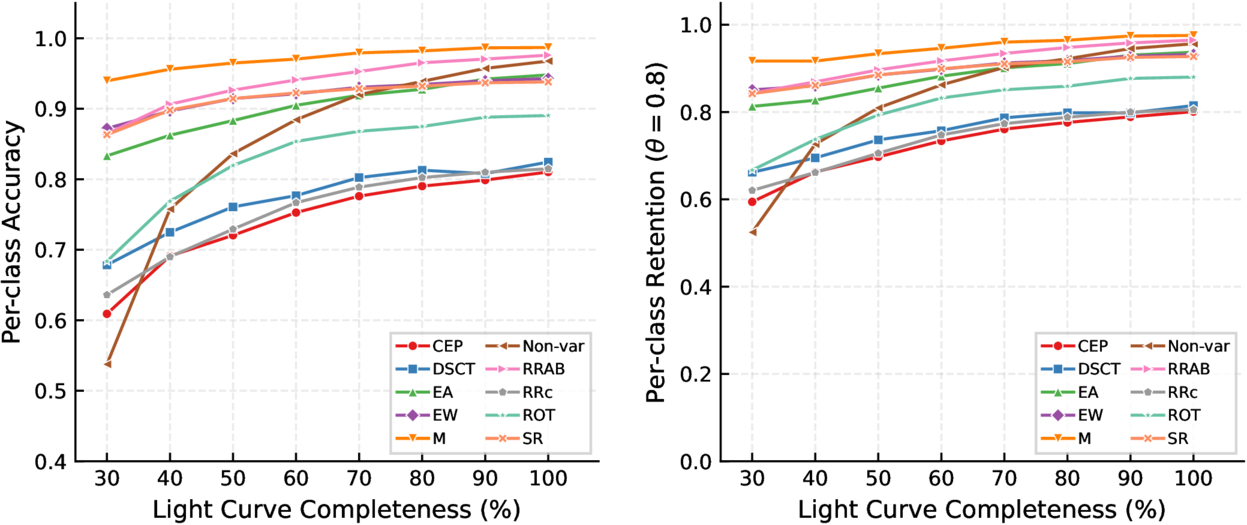}
\caption{Per-class metrics at fixed write-back threshold $\theta = 0.8$. \textbf{(a)}~Per-class classification accuracy vs.\ completeness. \textbf{(b)}~Per-class retention (fraction of sources in each class with $c \geq \theta$).}
\label{fig:cv_perclass}
\end{figure}

Although the classifier was trained on complete light curves, real-time applications often rely on partial observations \citep[e.g.,][]{muthukrishna2019rapid}. We therefore performed a retrospective truncation experiment using the same 5-fold splits as in Section~\ref{subsec:classification}. In each fold, models were trained on complete training curves, while test curves were truncated to $f \in \{30\%, 40\%, \dots, 100\%\}$ of full length. Features were re-extracted from the truncated test curves and evaluated with the fold-specific model. This is a controlled post hoc test of incomplete-curve performance, not a model explicitly trained for streaming classification.

For this truncation experiment, we sampled 1000 light curves from each class and applied seven truncation ratios, yielding about 70,000 truncated instances in total, which were evaluated under the standard 5-fold cross-validation inference setting. The overall accuracy increases monotonically with completeness, from 74.1\% at $f{=}30\%$ to 97.4\% at $f{=}100\%$, reaching 85.0\% at $f{=}50\%$. High-amplitude variables such as Mira stars (M, 92.6\% at $f{=}30\%$) can already be identified with high accuracy at low completeness, whereas lower-amplitude classes such as Non-var, ROT, and RRc generally require $\geq 60\%$ completeness before their predictions become consistently accurate (Figure~\ref{fig:cv_perclass}a).

\textbf{Confidence filtering.} To suppress unreliable predictions at early stages, we define a path-integrated confidence score for the hierarchical classifier. Because each source is routed through a sequence of sub-models, the confidence is computed as the product of the probabilities assigned to the selected branch at each decision layer. For example, a source classified as RRAB traverses four nodes (Variable $\to$ Intrinsic $\to$ RR $\to$ RRAB), and its confidence is
\begin{equation}
    c = \prod_{l=1}^{L} p_l,
\end{equation}
where $p_l$ denotes the probability assigned to the selected class at layer $l$, and $L$ is the length of the decision path. This score is used as a practical filtering criterion rather than as a calibrated posterior probability. Predictions with $c < \theta$ are withheld at the current stage and revisited after additional observations become available.

The confidence score in Eq.~(1) is the default path-level confidence used in \textsc{TDLight}. It is compared with a user-configurable threshold $\theta$ to control automatic database write-back. This path-integrated score is intentionally adopted as a conservative operating choice, because an apparently confident low-level classification may still be unreliable if an incorrect routing decision has already occurred at a higher level of the hierarchy. In this sense, the path product helps suppress premature write-back caused by upstream errors. At the same time, node-level conditional probabilities remain useful as diagnostic quantities and can be provided for branch-level inspection or user-defined operating policies. In practical use, the threshold $\theta$ can be adjusted according to the desired trade-off between retention and reliability.
\begin{figure}[t]
\centering
\includegraphics[width=\textwidth]{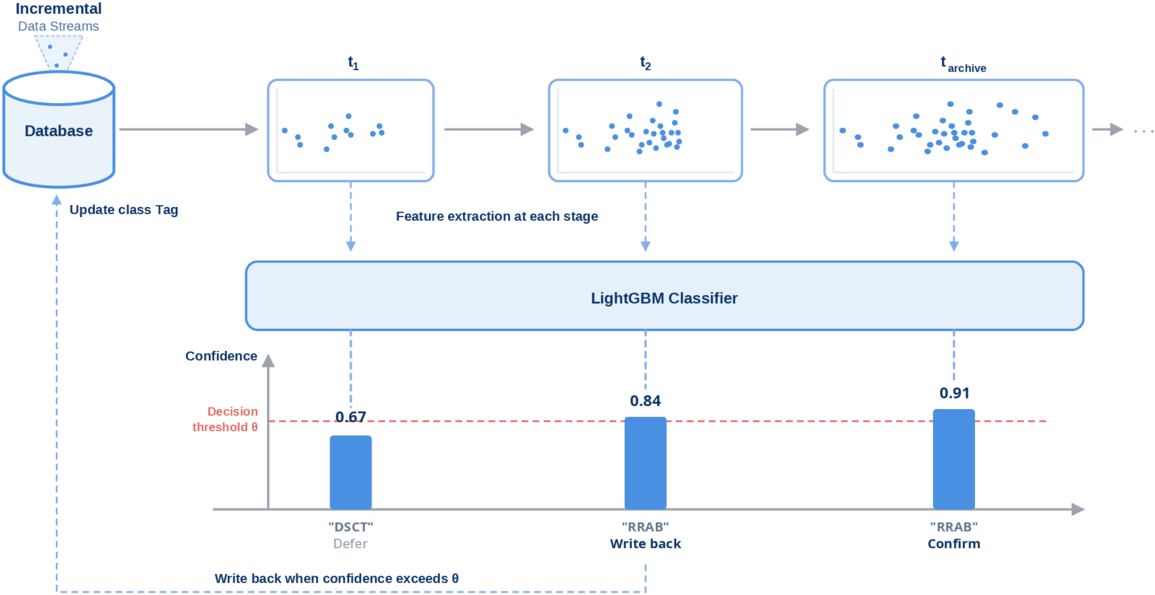}
\caption{Incremental classification and confidence-based database write-back in \textsc{TDLight}.}
\label{fig:class_mech}
\end{figure}

Figure~\ref{fig:cv_threshold} shows the accuracy--retention trade-off under different $\theta$: higher thresholds improve reliability at the cost of coverage.

\textbf{Trigger policy.} Classification is controlled by the confidence threshold $\theta$ and growth ratio $g$. Sources are selected if they are newly ingested and still unlabeled, or if their light curves have grown by more than $g$ since the previous classification (default $g{=}20\%$ of the previous length). For each selected source, the system extracts 15 features, applies the hierarchical classifier, and computes the path-integrated confidence score $c$. Only predictions with $c \geq \theta$ (default $\theta{=}0.8$) are written back to the database; otherwise, classification is deferred until more observations accumulate. Classification can also be triggered manually through the web interface.

\textbf{Design rationale.} The default $g{=}20\%$ balances responsiveness against computational cost: smaller values lead to frequent reclassification with limited feature changes, whereas larger values delay potentially improved predictions. The default $\theta{=}0.8$ is chosen as a practical operating point on the accuracy--retention curve (Figure~\ref{fig:cv_threshold}a), improving reliability while retaining most sources. Because early-stage light curves are incomplete and may differ substantially from the complete curves used for training, confidence filtering helps prevent unreliable write-back and enables more reliable early classification.

\subsection{Catalog Disagreements and Candidate Label Issues in \textsc{LEAVES}} \label{subsec:label_errors}

Beyond incremental classification, \textsc{TDLight} can also be used to identify potentially problematic catalog labels through high-confidence disagreements between model predictions and archived classifications. Among the 14,355 \textsc{LEAVES} sources successfully cross-matched with Simbad \citep{wenger2000simbad} and assigned prediction confidence $\geq 0.95$, 560 cases were identified in which the \textsc{TDLight} prediction disagrees with the original \textsc{LEAVES} label but is supported by the Simbad classification.

To further assess these candidates, we cross-matched them against additional external variability catalogs, including VSX \citep{watson2006vsx}, ASAS-SN \citep{kochanek2017asas,jayasinghe2018asas,jayasinghe2019asas,christy2023asas}, and GCVS \citep{samus2017gcvs}. Of the 560 candidates, 402 (71.8\%) are supported by at least two external catalogs in favor of the \textsc{TDLight} prediction, while the remaining 158 (28.2\%) show conflicting or insufficient evidence.

A particularly notable subset consists of 150 RRc$\rightarrow$EW disagreements among the 402 externally supported candidates. All 150 sources exhibit an exact 2:1 ratio between the periods listed in \textsc{LEAVES} and the independently reported periods in VSX and ASAS-SN. This systematic offset is consistent with the analysis of \citet{Jayasinghe2020}, who noted that some RRc variables in the ASAS-SN pipeline were assigned twice their true pulsation period and could consequently be confused with EW systems because of their relatively symmetric folded light-curve morphologies. Since the relevant \textsc{LEAVES} entries are derived from ASAS-SN, this pattern strongly suggests that the disagreement reflects an inherited period-doubling artifact rather than random model error, with the \textsc{TDLight} prediction recovering the physically more plausible RRc interpretation.

These results suggest that a substantial fraction of the observed disagreements are associated with systematic labeling issues in upstream survey products, rather than arbitrary classifier failures. \textsc{TDLight} therefore provides value not only as an incremental classification framework, but also as a practical tool for catalog auditing by highlighting candidates for label verification, period re-analysis, and follow-up inspection. The source-by-source catalog of the 560 high-confidence disagreement candidates identified in this work is available on Zenodo: \dataset[doi:10.5281/zenodo.19736413]{https://doi.org/10.5281/zenodo.19736413} and summarized in Table \ref{mrt:descript}.

\begin{deluxetable*}{llll}
\tablecaption{Source-by-source catalog of high-confidence disagreement candidates \label{mrt:descript}}
\tablehead{
   \colhead{Column \#} & \colhead{Units} & \colhead{Label} & \colhead{Description}
}
\startdata
1 &   ---   & Index         & Original input-catalog row index [row\_index] \\
2 &   ---   & Name          & Source name [source\_name] \\
3 &   ---   & True          & Original input-catalog class label [true\_label] \\
4 &   ---   & Pred          & TDLight predicted class label [pred\_label] \\
5 &   ---   & Conf          & TDLight path confidence [confidence] \\
6 &   d     & Per           & TDLight Lomb-Scargle period [period] \\
7 &   mag   & Amp           & Peak-to-peak amplitude [amplitude] \\
8 &   deg   & RAdeg         & Right ascension (J2000) [ra\_deg] \\
9 &   deg   & DEdeg         & Declination (J2000) [dec\_deg] \\
10 &   ---   & Simbad        & SIMBAD identifier [simbad\_id] \\
11 &   ---   & otype         & SIMBAD object type [simbad\_otype] \\
12 &   ---   & VSX           & VSX name [vsx\_name] \\
13 &   ---   & Type-VSX      & VSX object type [vsx\_type] \\
14 &   d     & Per-VSX       & VSX period [vsx\_period] \\
15 &   ---   & ASASSN        & ASAS-SN name [asassn\_name] \\
16 &   ---   & Type-ASASSN   & ASAS-SN object type [asassn\_type] \\
17 &   d     & Per-ASASSN    & ASAS-SN period [asassn\_period] \\
18 &   mag   & Amp-ASASSN    & ASAS-SN amplitude [asassn\_amp] \\
19 &   ---   & GCVS          & GCVS name [gcvs\_name] \\
20 &   ---   & Type-GCVS     & GCVS object type [gcvs\_type] \\
21 &   d     & Per-GCVS      & GCVS period [gcvs\_period] \\
22 &   ---   & Pred-broad    & Predicted broad class [pred\_broad] \\
23 &   ---   & True-broad    & Broad class of original label [true\_broad] \\
24 &   ---   & VSX-broad     & Broad class from VSX type [vsx\_broad] \\
25 &   ---   & ASASSN-broad  & Broad class from ASAS-SN type [asassn\_broad] \\
26 &   ---   & Simbad-broad  & Broad class from SIMBAD type [simbad\_broad] \\
27 &   ---   & Ncpred        & Catalog count supporting predicted broad class [n\_confirm\_pred] \\
28 &   ---   & Nctrue        & Catalog count supporting original broad class [n\_confirm\_true] \\
29 &   ---   & Final         & Final cross-catalog assessment [final\_verdict] \\
\enddata
\tablecomments{The labels in square brackets in this descriptive table correspond to the labels used for the data file available on Zenodo.}
\tablecomments{Broad-class labels merge detailed types for cross-catalog comparison (e.g., EA/EW $\rightarrow$ EB; RRAB/RRc $\rightarrow$ RR; SR/M $\rightarrow$ LPV). MODEL\_LIKELY\_CORRECT means Ncpred $>$ Nctrue; otherwise CONFLICTING.}
\tablecomments{Per is the \textsc{TDLight} Lomb-Scargle feature value and may differ from external-catalog periods, especially for some long-period sources.}
\tablecomments{Table \ref{mrt:descript} is published in its entirety in the electronic edition of the {\it Astrophysical Journal}.  A portion is shown here for guidance regarding its form and content.}
\end{deluxetable*}

\section{Deployment and Interface} \label{sec:deployment}
\begin{figure}[t]
    \centering
    \includegraphics[width=\textwidth]{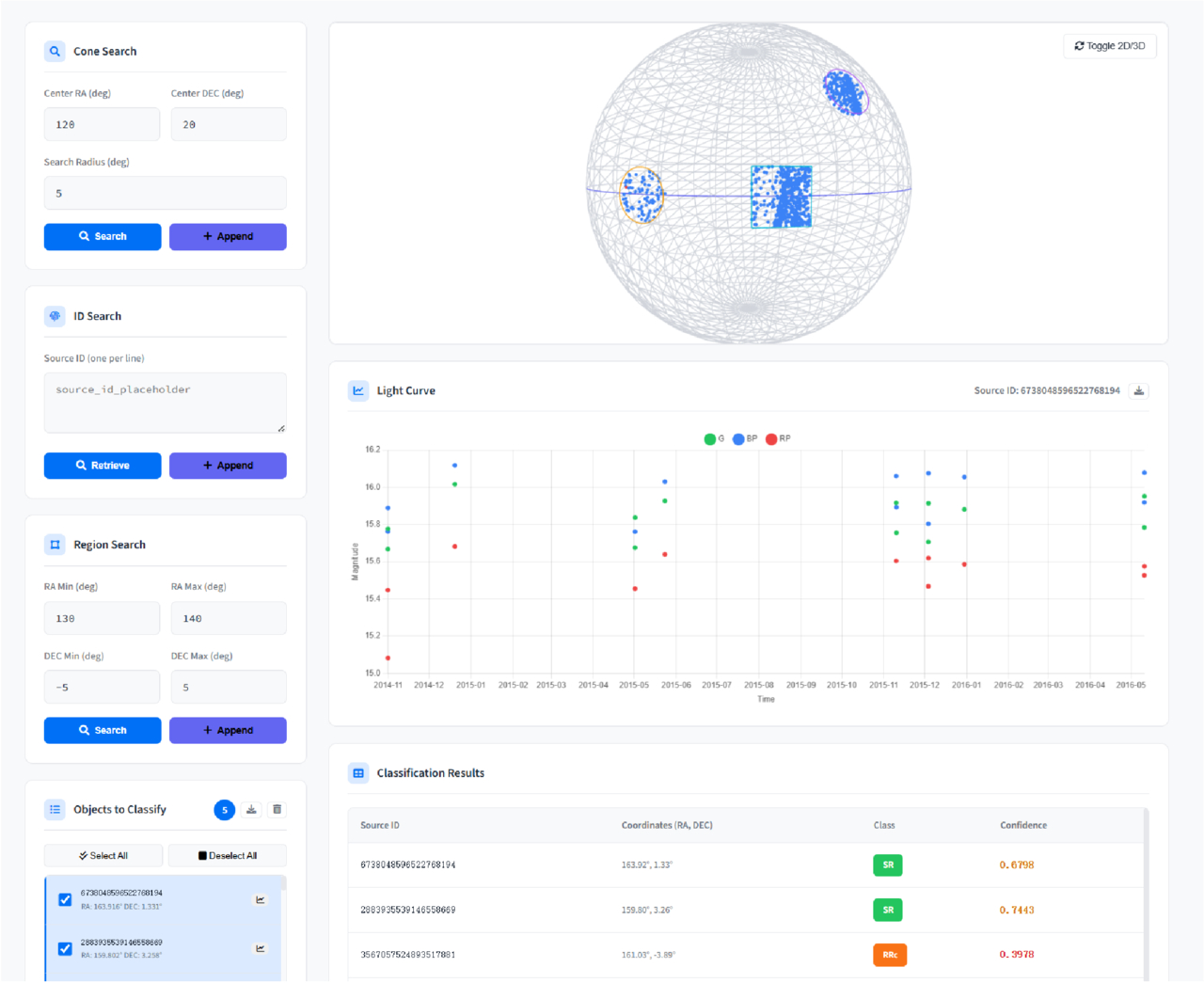}
    \caption{The \textsc{TDLight} web interface. The left panel shows spatial querying; the right panel shows light curves and classification metadata.}
    \label{fig:web_interface}
    \end{figure}
\subsection{Open-Source Release and One-Click Installation}
\textsc{TDLight} is released as an open-source project on \href{https://github.com/bestdo77/TD-light}{GitHub}. An automated installation script (\texttt{install.sh}) sets up \textsc{TDengine}, compiles the C++ ingestion and query modules, configures HEALPix, and deploys the web server and classification engine. The pre-trained LightGBM model is hosted on \href{https://huggingface.co/bestdo77/Lightcurve_lgbm_111w_15_model}{Hugging Face} and can be integrated directly after installation.

\subsection{Web Interface}
\textsc{TDLight} provides a browser-based frontend for cone and rectangular spatial queries, source lookup, light-curve inspection, and on-demand reclassification. A built-in data-import panel allows uploading light-curve archives and epoch catalogs through the GUI, with real-time progress monitoring via Server-Sent Events (Figure~\ref{fig:web_interface}).

\subsection{Interactive Label Correction and Incremental Model Update}

In addition to automatic trigger-based classification, \textsc{TDLight} also supports user-guided model refinement through the web interface. Users may upload source-level light-curve files in CSV or ZIP format, optionally providing corrected labels through a \texttt{class}, \texttt{label}, or \texttt{type} field. The system then extracts the same 15 statistical features used by the online classifier and stores them as incremental training samples for the corresponding branches of the hierarchical model.

Rather than retraining the full classifier from scratch, \textsc{TDLight} updates only the affected LightGBM sub-models in the four-level hierarchy. This design preserves the lightweight deployment target of the system while enabling rapid incorporation of user feedback. The updated models can then be re-exported to ONNX format and reused by the online classification pipeline.

To improve robustness in practical use, the incremental update procedure includes several safety mechanisms, including automatic model backup before training, rollback on failure, and deferred ONNX export in the background. This design extends confidence-based database write-back to a human-in-the-loop workflow, where expert corrections can be incorporated into the deployed classifier without interrupting the surrounding storage and query services.

\section{Conclusions and Outlook} \label{sec:conclusion}

We have presented \textsc{TDLight}, a database-centered framework that integrates incremental light-curve ingestion, repeated classification, and confidence-based database write-back within a single workflow. On a Gaia-derived workload, the system achieves archival ingestion throughput of ${\sim}1.5\times10^{6}$~rows\,s$^{-1}$ and supports sub-second spatial queries on a single server without GPU support. On the \textsc{LEAVES} benchmark, the hierarchical LightGBM classifier reaches an overall accuracy of 93.6\%. Because classification is re-triggered as light curves grow, the confidence filter allows reliable predictions to be accepted before the full observing campaign is complete. Beyond online classification, \textsc{TDLight} also identifies 560 high-confidence catalog-disagreement candidates, indicating its potential utility for catalog verification and follow-up analysis.

Two directions for future work are particularly promising. First, \textsc{TDengine} supports distributed deployment across multiple nodes, and extending the current single-server implementation to a distributed setting would enable the same workflow to scale to much larger time-domain streams. Second, the classification module is designed to be replaceable: more expressive models for irregular and incrementally growing light curves, including Transformer-based approaches, could be incorporated without changing the surrounding ingestion and storage pipeline. These extensions would further improve the applicability of \textsc{TDLight} to large-scale time-domain surveys.

\begin{acknowledgments}

This work was supported by the National Natural Science Foundation of China (NSFC; 12273025, 12133010, 62402333). Data resources are supported by the China National Astronomical Data Center (NADC) and the Chinese Virtual Observatory (China-VO). We also acknowledge \textsc{TDengine} as the open-source time-series database that underpins this work.

We thank the authors of the \textsc{LEAVES} project \citep{fei2024leaves} for publicly releasing their dataset and hierarchical classification resources, which enabled the evaluation and comparison presented in this work. We thank Las Cumbres Observatory and its staff for their continued support of ASAS-SN. ASAS-SN is funded in part by the Gordon and Betty Moore Foundation through grants GBMF5490 and GBMF10501 to the Ohio State University, and in part by the Alfred P. Sloan Foundation through grant G-2021-14192.

This work has made use of data from the European Space Agency (ESA) mission Gaia (\url{https://www.cosmos.esa.int/gaia}), processed by the Gaia Data Processing and Analysis Consortium (DPAC, \url{https://www.cosmos.esa.int/web/gaia/dpac/consortium}). Funding for the DPAC has been provided by national institutions, in particular the institutions participating in the Gaia Multilateral Agreement. This work also makes use of observations obtained with the Samuel Oschin 48-inch Telescope and the 60-inch Telescope at the Palomar Observatory as part of the Zwicky Transient Facility project. ZTF is supported by the National Science Foundation under grant Nos. AST-1440341 and AST-2034437, and by a collaboration including Caltech, IPAC, the Oskar Klein Center at Stockholm University, the University of Maryland, the University of California, Berkeley, the University of Wisconsin at Milwaukee, the University of Warwick, Ruhr University, Cornell University, Northwestern University, and Drexel University. Operations are conducted by COO, IPAC, and UW.

\end{acknowledgments}

\appendix
\section{Classification Module Interface}
\label{app:classification-interface}

The classification module in TDLight is decoupled from the database, ingestion trigger, and web interface through explicit input and output interfaces. In the released implementation (GitHub: \url{https://github.com/bestdo77/TD-light}), \texttt{class/classify\_pipeline.py} and \texttt{class/auto\_classify.py} handle database I/O and write-back, \texttt{class/hierarchical\_predictor.py} implements the model adapter, and \texttt{web/app.js} reads the classification results through REST endpoints defined in \texttt{web/web\_api.cpp}. A replacement classifier can therefore be integrated by preserving the model-side interfaces summarized in Table~\ref{tab:classification_interface}.

\begin{table}[ht]
\caption{\textbf{Model-side interfaces around the TDLight classification module.\label{tab:classification_interface}}}
\centering
\small
\renewcommand{\arraystretch}{1.12}
\setlength{\tabcolsep}{6pt}
\begin{tabular}{lll}
\hline\hline
\textbf{Interface} & \textbf{Type} & \textbf{API and content} \\
\hline
\textbf{Input} & \textbf{API:} & \texttt{classify\_pipeline.py, auto\_classify.py:} \texttt{fetch\_lightcurve()}\\
 & \textbf{Content:} & retrieves \texttt{ts}, \texttt{mag}, and \texttt{mag\_error} as \texttt{(t, mag, err)}\\
 & \textbf{API:} & \texttt{classify\_pipeline.py:} \texttt{extract\_features()}\\
 & \textbf{Content:} & converts light curves into the 15-dimensional feature vector used by the model\\
\textbf{Output} & \textbf{API:} & \texttt{hierarchical\_predictor.py:} \texttt{predict()}\\
 & \textbf{Content:} & returns the predicted class label and confidence score\\
 & \textbf{API:} & \texttt{web\_api.cpp: /api/classify\_results}\\
 & \textbf{Content:} & exposes result records with \texttt{source\_id}, \texttt{prediction}, and \texttt{confidence}\\
 & \textbf{API:} & \texttt{app.js:} \texttt{displayResults()}\\
 & \textbf{Content:} & displays classification results and optional metadata fields\\
\hline
\end{tabular}
\end{table}

Under this interface, a replacement module may use a different light-curve encoder or classifier, provided that it accepts the source-level light-curve or feature inputs supplied by the pipeline and returns the result fields consumed by the frontend and threshold-controlled database write-back logic. In the current implementation, database updates are performed by \texttt{classify\_pipeline.py: update\_tdengine\_class()} and \texttt{auto\_classify.py: update\_class\_in\_db()}, which write the predicted label back to TDengine only when the returned score exceeds $\theta$.

\software{
    TDengine v3.4,
    HEALPix \citep{gorski2005healpix},
    feets v0.4 \citep{cabral2018feets},
    LightGBM \citep{ke2017lightgbm},
    ONNX Runtime v1.17,
    Python v3.9,
    scikit-learn v1.1 \citep{JMLR:v12:pedregosa11a}
}

\clearpage
\bibliographystyle{aasjournalv7}
\bibliography{new.ms}

\end{document}